\documentclass[%
 reprint,
superscriptaddress,
nofootinbib,
 amsmath,amssymb,
 aps,
prd,
floatfix,
]{revtex4-2}

\usepackage{natbib}
\usepackage{orcidlink}
\usepackage{import}
\usepackage{amsmath}
\usepackage{amssymb}
\usepackage{hyperref}
\hypersetup{
    colorlinks=true,
    linkcolor=blue,
    filecolor=magenta,      
    urlcolor=blue,
    citecolor=violet
}

\usepackage{graphicx}
\usepackage{dcolumn}
\usepackage{bm}

\newcommand{\blip}{{\tt BLIP} }
\newcommand{\blips}{{\tt BLIP}'s\ }

\newcommand{\ed}[1]{#1}

\newcommand{\bgal}{{\tt B20} }
\newcommand{\popmap}{{\tt popmap} }

\begin{document}

\preprint{APS/123-QED}

\title{Templated Anisotropic Analyses of the LISA Galactic Foreground}

\author{Alexander W. Criswell\,\orcidlink{0000-0002-9225-7756}}
\affiliation{Department of Physics and Astronomy, Vanderbilt University, Nashville, TN 37240}
\affiliation{Department of Life and Physical Sciences, Fisk University, Nashville, TN 37208}
\affiliation{Minnesota Institute for Astrophysics, University of Minnesota, Minneapolis, MN 55455, USA}
\affiliation{School of Physics and Astronomy, University of Minnesota, Minneapolis, MN 55455, USA}

\author{Steven Rieck\,\orcidlink{0009-0006-0978-7892}}
\affiliation{School of Physics and Astronomy, University of Minnesota, Minneapolis, MN 55455, USA}
\affiliation{Department of Physics, University of Cincinnati, Cincinnati, OH 45221, USA}

\author{Vuk Mandic\,\orcidlink{0000-0001-6333-8621}}
\affiliation{Minnesota Institute for Astrophysics, University of Minnesota, Minneapolis, MN 55455, USA}
\affiliation{School of Physics and Astronomy, University of Minnesota, Minneapolis, MN 55455, USA}





\date{\today}

\begin{abstract}
The Laser Interferometer Space Antenna (LISA) will feature a prominent anisotropic astrophysical stochastic gravitational wave signal, arising from the tens of millions of unresolved mHz white dwarf binaries in the Milky Way: the Galactic foreground. While proper characterization of the Galactic foreground as a noise source will be crucial for every LISA science goal, it is extremely scientifically interesting in its own right, comprising --- along with $\sim10^4$ resolvable white dwarf binaries --- a complete sample of every mHz white dwarf binary in our Galaxy. We present a novel Bayesian analysis of the LISA Galactic foreground that directly treats its anisotropy via astrophysically-motivated templates, allowing for a direct connection between the observed time-modulation of the foreground amplitude and the underlying spatial distribution of the Milky Way. We validate the efficacy of this approach via simulated data and show that it is able to accurately recover the foreground spectrum in the presence of LISA instrumental noise.
\end{abstract}

\maketitle

\section{Introduction}\label{sec:intro}

\subsection{The LISA Galactic Foreground}\label{sec:intro_foreground}
The Laser Interferometer Space Antenna (LISA) \citep{amaro-seoane_laser_2017} is a spaceborne gravitational-wave (GW) observatory with a planned launch date in 2035. LISA will observe a wide variety of GW sources across the mHz band, including double white dwarf binaries (DWDs) \citep{nelemans_the_2001,ruiter_the_2010,korol_prospects_2017}, massive binary black hole mergers \citep{sesana_the_2005}, stellar-origin binaries far from merger \citep{sesana_prospects_2016,cutler_what_2019}, extreme mass ratio inspirals \citep{hils_gradual_1995, sigurdsson_capture_1997}, and more; see \citet{amaro-seoane_astrophysics_2023} for a review. Some of these sources are expected to be individually resolvable; however, others will be present within LISA's datastream in sufficient number and density that their individual GW contributions will be indistinguishable from one another, forming several stochastic gravitational wave backgrounds (SGWBs). Signals of this nature have been searched for above the LISA frequency range by the LIGO-Virgo-KAGRA (LVK; \citep{aasi_advanced_2015,acernese_advanced_2015,akutsu_kagra:_2019}) collaborations \citep{abbott_search_2019a, abbott_search_2021, abbott_upper_2017, abbott_upper_2021}. Notably, pulsar timing arrays have recently announced evidence of a nHz SGWB \citep{agazie_the_2023,agazie_the_2023a,agazie_the_2023e,antoniadis_the_2023a,antoniadis_the_2023,reardon_the_2023,reardon_search_2023,zic_the_2023,xu_searching_2023}. LISA data is expected to contain SGWBs from Galactic DWDs \citep{nelemans_the_2001,edlund_the_2005,ruiter_the_2010}, DWDs in the Large Magellanic Cloud \citep{rieck_a_2024}, and stellar-origin black hole binaries \citep{babak_stochastic_2023,chen_stochastic_2019, cusin_properties_2019a, perigois_startrack_2021,abbott_upper_2021,lewicki_impact_2023}. Depending on the rate of extreme mass ratio inspirals, LISA may also observe a SGWB from these sources, although the amplitude and characteristics of this SGWB are currently uncertain \citep{bonetti_gravitational_2020,pozzoli_computation_2023e,naoz_the_2023a}. Additionally, LISA has the potential to observe a number of theorized SGWBs of cosmological origin; see \citet{caprini_cosmological_2018}, \citet{auclair_cosmology_2023} for a review.

The Galactic foreground produced by the tens of millions of unresolved mHz DWDs in the Milky Way (MW) will likely be the most prominent SGWB present within LISA \citep{nelemans_the_2001,edlund_the_2005,ruiter_the_2010}. This signal will be highly anisotropic --- tracing the shape of the Galactic bulge and plane --- and so loud as to sit above the LISA instrumental noise (hence, ``foreground" as opposed to ``background"). Due to its foreground status, accurate characterization of the Galactic foreground will be crucial for every facet of LISA data analysis, as it will serve as a significant source of noise for analyses of not just other SGWBs, but also LISA's individually-resolvable signals \citep{nelemans_the_2001,edlund_the_2005}. Beyond its implications for other analyses, however, the Galactic foreground is extremely astrophysically interesting in its own right as --- in conjunction with the few tens of thousands of resolved DWD systems --- it represents a complete sample of all mHz DWDs in our Galaxy \citep{lamberts_predicting_2019}. Studying the Galactic foreground can provide us with insight into many aspects of the MW, including its morphology \citep{breivik_constraining_2020}, metallicity \citep{thiele_applying_2021}, star formation history \citep{yu_the_2013}, and more. \ed{Taken together, these factors mandate the development of comprehensive Galactic foreground analyses, which not only accurately characterize the Galactic foreground spectrum so as to enable the rest of LISA's science goals, but also treat the Galactic foreground as the anisotropic astrophysical signal it is --- thereby providing a new window into the dynamics and history of our Galaxy. In this work, we take the first step towards realization of this goal: a MW foreground analysis which leverages fixed spatial templates to directly connect the foreground anisotropy to the underlying astrophysical distribution of its contingent sources.}

\subsection{Galactic Foreground Analyses}\label{sec:intro_analyses}

\subsubsection{Foreground Analyses in Context: The Global Fit}
In order to characterize the wide array of source types and signal morphologies observable by LISA --- all of which are expected to be simultaneously present within its datastream --- a ``Global Fit" data analysis framework for LISA has been proposed. Such approaches seek to fit all noise and GW contributions to the data simultaneously, inferring individual source parameters as well as the overall number and classification of sources in a Bayesian fashion. Several prototype Global Fits currently exist, including GLASS \citep{littenberg_global_2020,littenberg_prototype_2023} and Erebor \citep{katz_an_2024a}.

At time of writing, no extant Global Fit prototype has incorporated detailed SGWB analyses\footnote{Although a prototype analysis has been developed to search for an isotropic SGWB in Global Fit residuals \citep{rosati_prototype_2024}} nor do they treat the Galactic foreground as the anisotropic astrophysical SGWB it is. Both Erebor and GLASS are currently limited to resolved Galactic binaries, massive black hole binary mergers, and simple models of the Galactic foreground. These simple foreground models forego full treatment of the Galactic foreground's anisotropy, instead either treating its power spectral density (PSD) as a time-independent noise contribution \citep{katz_an_2024a} or representing the effects of the Galactic foreground's interaction with the time-dependent directional response of the LISA detector in terms of a parameterized quasiperiodic amplitude modulation \citep[e.g.,][]{digman_lisa_2022,littenberg_prototype_2023}.

\subsubsection{Other Foreground Analyses}\label{sec:intro_asgwbs_other}
However, more advanced analyses of the Galactic foreground and other SGWBs in LISA have been developed; these include several frequentist map-making approaches \citep{ungarelli_studying_2001,kudoh_probing_2005,taruya_probing_2005,renzini_mapping_2018,contaldi_maximum_2020,bartolo_probing_2022}, which treat anisotropies but are not intrinsically compatible with the Bayesian Global Fit infrastructure (as they are not themselves Bayesian in nature). Conversely, \citet{boileau_spectral_2021} presented a Bayesian analysis for investigating SGWBs in LISA, but did not consider anisotropies, instead assessing the time-modulation of the Galactic foreground by independently inferring its amplitude in each of a series of individual time segments of LISA data. \citet{flauger_improved_2021} also developed a Bayesian foreground analysis, but neglected the anisotropy-induced time modulation entirely. Such approaches, while in principle suitable for integration into a Bayesian Global Fit, fail to capture information contained within the correlation of the MW foreground signal across time segments --- information which is accessible through detailed treatment of the Galactic foreground anisotropy. While all these approaches possess their individual strengths, the fact remains that optimal characterization of the Galactic foreground in a Global Fit setting requires an analysis that is both Bayesian and anisotropic in nature. \ed{Bayesian ASGWB analyses have been developed for terrestrial detectors \citep[e.g.,][]{chung_untargeted_2023,tsukada_bayesian_2023} and pulsar timing arrays \citep[e.g.,][]{gair_mapping_2014, taylor_from_2020a, taylor_searching_2013b}; however no such analysis existed for LISA before the advent of the Bayesian LISA Inference Package ({\tt BLIP}) \citep{banagiri_mapping_2021}. \blip has recently been joined in this capacity by the analysis of \citet{pozzoli_cyclostationary_2024}, which directly connects the cyclostationarity of the observed foreground modulation with its underlying anisotropy, and that of \citet{buscicchio_a_2024a} and \citet{piarulli_a_2024b}, which treats the effects of the foreground anisotropy through direct integration of a simplified analytic model.}


\subsubsection{\blip}\label{sec:intro_asgwbs_blip}
{\tt BLIP}, as presented in \citet{banagiri_mapping_2021}, is capable of a spherical harmonic search for anisotropic SGWBs (ASGWBs) that infers the coefficients of a spherical harmonic expansion of the square root of the ASGWB power on the sky. This approach mathematically ensures that the inferred spatial distribution is real and non-negative at every point in the sky; see \citet{banagiri_mapping_2021} for further details. The \blip spherical harmonic search has been previously applied to anisotropic analyses of a simple model of the Galactic foreground \citep{banagiri_mapping_2021}, as well as the ASGWB arising from unresolved DWD systems in the Large Magellanic Cloud \citep{rieck_a_2024}.

While the \blip spherical harmonic basis ASGWB analysis is well-suited to inferring \textit{a priori} unknown spatial distributions, it is a less effective choice for ASGWB models where the specific realization of the anisotropy is known. This is due to two factors: first, that the fidelity of the spherical harmonic basis to the true spatial distribution is reduced at computationally-accessible values of the spherical harmonic series truncation value $\ell^a_{\mathrm{max}}$; and second, that the spherical harmonic search cannot effectively leverage prior knowledge of such an ASGWB's spatial distribution, and as such must explore a much broader parameter space than a more constrained model. In contrast, the spatial distribution of Galactic DWDs is expected to be well-measured by LISA observations of its $\sim10,000$ individually resolvable DWD systems \citep[e.g.,][]{littenberg_prototype_2023}. This information could prove to be a powerful tool for accurate characterization of the Galactic foreground through targeted treatment of its anisotropy.

We therefore present a Bayesian Galactic foreground analysis for LISA within the \blip framework that leverages pixel-basis templates for the MW DWD population's anisotropy to enable a targeted anisotropic search for the Galactic foreground. The pixel basis approach is developed in \S\ref{sec:pixel_basis}; its application to templated analyses of the Galactic foreground is described in \S\ref{sec:pixel_templates}. Details of the spectral models used in this work are presented in \S\ref{sec:foreground_spec} and \S\ref{sec:noise}. Our data simulation procedure is discussed in \S\ref{sec:sims}. The results of our templated analysis as applied to simulated LISA data are presented in \S\ref{sec:results_simple} and \S\ref{sec:results_population} for a simplified model of the MW and a realistic, population-derived Galactic foreground, respectively. Finally, we discuss in \S\ref{sec:conclusion} the implications and utility of such an analysis, as well as directions of future work.

\section{Methods}\label{sec:methods}

\subsection{LISA Response Functions in the Pixel Basis}\label{sec:pixel_basis}
We present here in brief the underlying formalism of \blip and discuss its extension to analyses in the pixel basis, following derivations in \citep{cornish_detecting_2001,cornish_space_2001,adams_discriminating_2010,romano_detection_2017,banagiri_mapping_2021}. For further details on the \blip formalism beyond what is discussed herein, refer to \citet{banagiri_mapping_2021}. 

\blip \ed{performs SGWB analyses by dividing the LISA datastream into segments of length $T_{\mathrm{seg}}=10^5$ s, and taking the short-time Fourier transform of each segment. The value of $T_{\mathrm{seg}}$ was chosen to exceed the autocorrellation timescale of LISA data in this frequency band ($\sim10^4$~s), while ensuring that the adiabatic approximation holds for the satellite's motion \cite{banagiri_mapping_2021}. It then} uses a variant of the multi-dimensional complex Gaussian likelihood \citep{adams_discriminating_2010} in terms of \ed{the piece-wise stationary} time- and frequency-dependent covariance matrix $C_{IJ}(\ed{f,t})$, where $t$ indexes the $t^{th}$ time segment and $f$ the $f^{th}$ frequency bin:
\begin{widetext}
\begin{equation}\label{eq:blip_likelihood}
    \mathcal{L}(d|\vec\theta) = \prod_{f,t}\frac{1}{2\pi T_{\mathrm{seg}}\det\left(C_{IJ}(\ed{f,t}|\vec\theta)\right)}
    \times\exp\left[-\frac{2}{T_{\mathrm{seg}}}\sum_{I,J}\tilde{d}_I^*(f,t)\left(C(\ed{f,t}|\vec\theta)\right)^{-1}_{IJ}\tilde{d}_J(f,t)\right],
\end{equation}
\end{widetext}
where $\tilde{d}_I$ is the Fourier-domain data in channel I, and $\{I,J\}$ index over the three LISA Time Domain Interferometry (TDI) channels \citep{tinto_time-delay_2020}.
As this work considers a rigid, equal-armed LISA constellation, we use the XYZ TDI channels for all results, although the analysis can be straightforwardly extended to the unequal-arm case with the 2\textsuperscript{nd}-generation AET TDI channels \citep{tinto_second-generation_2023}. \ed{This formulation necessarily assumes independence across time-segments, thereby allowing us to multiply the individual Whittle likelihoods of all segments together to arrive at the full joint likelihood.} The entries of the covariance matrix $C_{IJ}(\ed{f,t})$ are defined such that
\begin{equation}
    C_{IJ}(\ed{f,t}) = S^n_{IJ}(f) + S^{\mathrm{GW}}_{IJ}(f,t),
\end{equation}
where $S^n_{IJ}$ and $S^{\mathrm{GW}}_{IJ}$ are the noise and GW auto/cross-correlation PSDs for channels $I$ and $J$, respectively. $S^{\mathrm{GW}}_{IJ}$ is not, however, the true astrophysical SGWB PSD; this latter quantity is denoted $S_{\mathrm{GW}}(f,\ed{\mathbf{n}})$, and is related to $S^{\mathrm{GW}}_{IJ}(f,t)$ through convolution with the LISA response functions $\mathcal{R}_{IJ}(f,t)$ (as we will see shortly). We assume that the ASGWB is separable into its spectral and spatial components:\footnote{\ed{While this assumption is generally accurate for extragalactic backgrounds, it is worth noting that the MW DWD population's mass and orbital separation distributions --- and therefore the shape of MW foreground spectrum --- may have some metallicity dependence \citep[e.g.,][]{thiele_applying_2021}. As the Milky Way's metallicity is likely spatially-dependent --- as seen for other spiral galaxies \citep[e.g.,][]{aller_the_1942,searle_evidence_1971,pagel_abundances_1981} --- there may be some low-level coupling between the Galactic DWD population's mass/orbital separation distributions and their spatial distribution. Such an effect would likely comprise a second- (or third-) order correction, so we will maintain the assumption of independence between these quantities for this work. One could envision, however, an analysis that seeks to leverage this connection to infer new insights into the spatial dependence of metallicity in the Milky Way; this is an intriguing possibility and should be explored in future.}}
\begin{equation}
    S_{\mathrm{GW}}(f,\ed{\mathbf{n}}) = S_{\mathrm{GW}}(f)\mathcal{P}(\ed{\mathbf{n}}),
\end{equation}
where we take the convention that $\mathcal{P}(\ed{\mathbf{n}})$ is normalized such that the integral of $\mathcal{P}(\ed{\mathbf{n}})$ over this sky is equal to 1, i.e.,
\begin{equation}\label{eq:blip_Pofn_norm}
    \int\mathcal{P}(\ed{\mathbf{n}})d^2\ed{\mathbf{n}} = 1.
\end{equation}
$S_{\mathrm{GW}}(f)$ can be recast in terms of the dimensionless GW energy density $\Omega_{\mathrm{GW}}$, such that
\begin{equation}\label{eq:omegaf_sgw}
    \Omega_{\mathrm{GW}}(f) = \frac{2\pi^2}{3H_0^2}f^3S_{\mathrm{GW}}(f).
\end{equation}

The LISA response functions to an incident (A)SGWB are known \citep{schilling_angular_1997,cornish_space_2001,cornish_detecting_2001}. They can be written in terms of the antenna pattern functions $F^A(f,t,\ed{\mathbf{n}})$ such that
\begin{equation}\label{eq:blip_response_fx_def}
    \mathcal{R}_{IJ}(f,t) = \frac{1}{2}\int\mathcal{P}(\ed{\mathbf{n}})\left(\sum_A F_I^A(f,t,\ed{\mathbf{n}})F_J^{A*}(f,t,\ed{\mathbf{n}})\right)d^2\ed{\mathbf{n}},
\end{equation}
where $A$ indexes over GW polarizations. \ed{For convenience, we will also define the full time, frequency, and directionally-dependent LISA response functions $\mathcal{R}(f,t,\ed{\mathbf{n}})$ as
\begin{equation}\label{eq:blip_response_fx_ftn}
    \mathcal{R}_{IJ}(f,t,\ed{\mathbf{n}}) = \frac{1}{2}\left(\sum_A F_I^A(f,t,\ed{\mathbf{n}})F_J^{A*}(f,t,\ed{\mathbf{n}})\right).
\end{equation}}
The antenna pattern functions encode the LISA constellation geometry, the timing transfer functions for an incident GW on an interferometric detector and so on; see \citet{romano_detection_2017} for a derivation of the general case, \citet{schilling_angular_1997,cornish_space_2001,cornish_detecting_2001} for LISA in particular, and \citet{banagiri_mapping_2021} for these functions as implemented in \blip. The full ASGWB cross-correlation PSD can then be written as
\begin{widetext}
    \begin{equation}\label{eq:sgw_response_fx}
    \begin{split}
        S^{\mathrm{GW}}_{IJ}(f,t) &= \frac{1}{2}\int S_{\mathrm{GW}}(f,\ed{\mathbf{n}})\left(\sum_A F_I^A(f,t,\ed{\mathbf{n}})F_J^{A*}(f,t,\ed{\mathbf{n}})\right)d^2\ed{\mathbf{n}}\\
        &=\frac{1}{2}\int\mathcal{P}(\ed{\mathbf{n}})\left(\sum_A F_I^A(f,t,\ed{\mathbf{n}})F_J^{A*}(f,t,\ed{\mathbf{n}})\right)d^2\ed{\mathbf{n}} \times S_{\mathrm{GW}}(f)\\
        &\ed{=\int\mathcal{P}(\ed{\mathbf{n}})\,\mathcal{R}_{IJ}(f,t,\ed{\mathbf{n}})\,d^2\ed{\mathbf{n}} \times S_{\mathrm{GW}}(f)}\\
        & = \mathcal{R}_{IJ}(f,t)\times S_{\mathrm{GW}}(f).
    \end{split}
    \end{equation}
\end{widetext}

The \blip pixel-basis ASGWB analysis presented in this work explicitly represents $\mathcal{P}(\ed{\mathbf{n}})$ as a Healpix \citep{gorski_healpix:_2005} skymap via Healpy \citep{zonca_healpy:_2019}, where the amplitude of each pixel corresponds to the area-averaged integrated GW power contained within it, subject to the usual normalization of $\mathcal{P}(\ed{\mathbf{n}})$ per Eq.~\eqref{eq:blip_Pofn_norm}. This pixel representation of the ASGWB spatial distribution is then convolved with the time-dependent directional LISA response functions $\mathcal{R}(f,t,\ed{\mathbf{n}})$ as computed over the same pixel grid. The integral in Eq.~\eqref{eq:sgw_response_fx} is approximated as a discrete sum over the pixelated skymap. Note that at minimum a Healpix {\tt nside} (skymap resolution) of 16 is needed to capture the spatial variation of the LISA directional response with sufficient resolution to avoid significant systematic error. This corresponds to pixel areas of 13.5 deg\textsuperscript{2}. Higher resolutions naturally further improve fidelity and the ability to model fine spatial detail; however, calculating the full LISA time-dependent directional response at higher resolutions incurs rapidly increasing computational costs, especially for data spanning the nominal LISA mission duration of 4 years. All results discussed in this work are computed using an {\tt nside} of 32.

\subsection{Pixel Basis Templates in \blip}\label{sec:pixel_templates}
In the pixel basis, one can model the spatial distribution of the Galactic foreground (or any ASGWB) in one of several ways: via an astrophysically parameterized model, a single fixed template, or a mix of the two which employs a discrete template bank spanning some portion of the relevant parameter space. This work focuses on the fixed template approach; astrophysically parameterized pixel-basis models of the Galactic foreground anisotropy are in development for \blip and will be presented in future. Extension of this work to a full template bank comprises a compelling direction of future research, but is beyond the scope of the current study. 

Using a fixed template is extremely advantageous from a computational perspective. Assuming a specific spatial distribution allows for full amortization prior to sampling of the (significant) computational cost of an astrophysically-motivated anisotropic Galactic foreground model. This includes simulation in 3D space of a model MW, computation of the corresponding spatial distribution of its GW power as seen by LISA, and convolution with the time-dependent directional response of the LISA detector. In a parameterized approach, this process must be performed at every iteration of the sampler, leading to intractably long sampling times.\footnote{(at present; this is, of course, ``simply" a problem of optimization.)} A fixed spatial template, however, allows us to perform this entire process only once, and reduces the overall 
computing time for each iteration of the sampler to roughly that of an isotropic analysis. 

Moreover, the memory (i.e., RAM) requirements for an anisotropic analysis are greatly reduced when employing a fixed template. Without a template, holding the full time-dependent directional LISA response in memory is the major limiting factor for \blips ability to analyze data with long durations/broad frequency ranges/high spatial resolution. This is due to the high dimensionality of the response array (i.e., $3\times3\;\times$ the number of time segments $\times$ the number of frequency bins $\times$ the number of sky pixels). For instance, a parameterized pixel-basis anisotropic analysis over the nominal LISA mission duration (4 years) with time segments of $10^5$ s, a frequency resolution of $\Delta f=0.01$ mHz on $f\in [0.1,10]$ mHz, and a Healpix {\tt nside} of 32 would require an array with (approximate) dimension of $3\times3\times1262\times10^3\times12288$, which for an array of complex double-precision floats yields a memory requirement of order 1-2 TB. In contrast, amortizing the spatial dependence drops this figure to order a few hundred MB to a few GB. Beyond the obvious advantages of not needing several TB of RAM to run one's analysis, the lightweight nature of the templated approach allows it to be run on GPUs, which have much lower RAM capability than distributed RAM on a CPU cluster (for instance, the Nvidia A100s used for the results presented in this study have 80 GB of available RAM).

The \blip implementation of templated anisotropic searches allows for any user-specified spatial distribution to be added as a template, provided it can be represented on a Healpy map. At present, \blip includes four distinct classes of templated search: a point source template that can search for ASGWB power from any number of user-specified sky locations; an extended compact source template suitable for MW satellite galaxies like the Large Magellanic Cloud; and the two Galactic foreground spatial templates considered in this work --- one based on an adjustable 2-parameter MW model following the simple MW model of \citet{mcmillan_mass_2011} and \citet{breivik_constraining_2020} (``{\tt B20}"), and a DWD population-derived template (``{\tt popmap}").

In principle, this templated anisotropic analysis is similar in conception to radiometer methods used for current terrestrial GW detectors \citep{ballmer_a_2006, mitra_gravitational_2008}. Such searches \citep[e.g.,][]{abbott_directional_2019,abbott_search_2021,agarwal_targeted_2022} assume that the target ASGWB consists of one or more discrete point sources, and allow the amplitude of each point source to vary independently. As many astrophysical ASGWBs ultimately arise from the superposition of many point sources, the utility of such an approach is immediately apparent. However, because of this assumption that each pixel is independent, radiometer searches are poorly-suited to analyses of diffuse ASGWBs due to the fact that they do not account for correlations between neighbouring pixels \citep{abbott_search_2021}. In the case of current-generation terrestrial detectors, lack of prior knowledge as to the spatial distribution of (e.g.) the SGWB from stellar-origin compact binary mergers lends motivation to the use of searches in the spherical harmonic basis (which can effectively handle correlations across the sky), whereas likely point source candidates such as the Galactic center merit a radiometer approach \citep[e.g.,][]{abbott_directional_2019,abbott_search_2021}. In contrast, LISA analyses of the Galactic foreground (and other LISA ASGWBs, such as that of the Large Magellanic Cloud \citep{rieck_a_2024}) have a distinct advantage: strongly informative prior knowledge as to the underlying astrophysical spatial distributions of their component binaries. The \blip pixel-basis ASGWB analysis allows us to directly leverage this prior knowledge to account for correlations across pixels by explicitly modelling for the spatial distribution of, e.g., the MW DWD population on the sky. Doing so allows us to maintain the intrinsic fidelity of the pixel basis while accurately characterizing diffuse-but-known ASGWBs. 

Templated analyses of the Galactic foreground in particular benefit from the fact that LISA is expected to measure the spatial distribution of resolvable Galactic DWD systems with excellent precision \citep[e.g.,][]{littenberg_global_2020}. Additionally, the sample of resolved DWDs within the Milky Way will not be distance-limited \citep{lamberts_predicting_2019}. Instead, the division between resolved and unresolved DWD systems will be primarily determined by each system's intrinsic GW amplitude and frequency --- and therefore by the overall Galactic DWD population distributions for mass, orbital separation, and inclination. An important consequence of this is that the spatial distributions of LISA's resolved and unresolved Galactic DWDs can be expected to be near-identical. This allows us to rely on the Galactic DWD spatial distribution inferred by analyses of the resolved binaries \citep[e.g.,][]{littenberg_prototype_2023} for analyses of their unresolved counterparts. For the purposes of this work, we assume we possess an accurate point-estimate of the resolved DWD spatial distribution and use this known spatial distribution as our fixed anisotropic template.

Note that this approach implicitly assumes that the ASGWB is itself stationary and that the observed time-modulation of (for instance) the Galactic foreground as seen in LISA's datastream is solely driven by the time-dependent directional response of the LISA detector as it sweeps over the static MW during the course of LISA's orbit. This avoids the need for ad hoc parameterized models of the Galactic foreground's (or any other ASGWB's) amplitude modulation over the course of LISA's orbit \citep[e.g.,][]{littenberg_prototype_2023} or independent modelling of the Galactic foreground amplitude within each of any number of individual time segments \citep[e.g.,][]{boileau_spectral_2021}. Moreover, this approach allows for explicit treatment of the connection between the underlying spatial distribution of MW DWDs and the observed time-dependence of the Galactic foreground in the LISA datastream, thereby enabling direct access to Galactic astrophysics of interest.

\subsection{Adjustable 2-Parameter MW Template ({\tt B20})}\label{sec:template_simple}
\bgal is a spatial template with two adjustable parameters, based on the simple MW model of \citet{mcmillan_mass_2011} and \citet{ breivik_constraining_2020} as implemented for simulation purposes in \citet{banagiri_mapping_2021}. This template directly models the overall MW DWD density distribution in 3D space as a cylindrically symmetrical bulge and exponential disk. Specifically, the DWD density in the disk is defined such that
\begin{equation}\label{eq:mw_template_disk}
    \rho_{\mathrm{disk}}(r,z) \propto \exp(-r/r_h)\exp(-z/z_h),
\end{equation}
and the bulge density as
\begin{equation}\label{eq:mw_template_bulge}
    \rho_{\mathrm{bulge}}(r,r') \propto \frac{\exp\left( 
-(r/r_{\mathrm{cut}})^2 \right)}{(1+r'/r_0)^{\gamma}},
\end{equation}
where $r' = \sqrt{r^2 + (z/q)^2}$, $q=0.5$, $\gamma=1.8$, $r_0=0.075$ kpc, and $r_{\mathrm{cut}}=2.1$ kpc. The radial and vertical scale height parameters, $r_h$ and $z_h$, can be adjusted to create the user's template of choice. This density distribution is computed on a 3D grid at a resolution of 0.33 kpc. The total GW strain contribution in LISA from each cell will be proportional to the number density of DWDs in the cell, modulated by the cell's distance from the Solar System Barycentre (SSB). As the template itself is normalized to 1 per Eq.~\eqref{eq:blip_Pofn_norm}, we are free to neglect the precise proportionality, and compute the template by integrating the distance-modulated density of all cells within every SSB-frame line of sight as represented on a Healpy pixelated skymap. This process yields an unnormalized map of the model MW's anisotropic GW strain distribution on the sky. The skymap is then masked to the first four scale heights --- i.e., 4 factors of $e$ in amplitude; for the exponential disk model of Eq.~\eqref{eq:mw_template_disk} this equates physically to radial and vertical disk scales in kpc of $4r_h$ and $4z_h$, respectively --- such that the minimum amplitude is approximately two orders of magnitude below the maximum. Doing so prevents inclusion of a nonphysical ``haze" extending off the Galactic disk into regions where the DWD density is greatly reduced and source discreteness becomes relevant, thereby invalidating our assumption that the DWD density can be used as a proxy for the stochastic GW power distribution on the sky. Masking the skymap in such a way also avoids the risk of biasing/overwhelming inference of an isotropic signal by allowing the Galactic foreground model to absorb diffuse SGWB power off the Galactic plane. The specific value of 4 scale heights used here was chosen to preserve the overall bulge+disk structure in the template. Following application of the mask, the skymap is normalized per Eq.~\eqref{eq:blip_Pofn_norm} to produce the template $\mathcal{P}(\ed{\mathbf{n}})$, which can then be paired with any desired spectral model.

\subsection{Population-derived Template ({\tt popmap})}\label{sec:template_pop}
The second MW foreground spatial template, {\tt popmap}, is designed to capture the spatial distribution of a realistic Galactic foreground. It is an extension of \blips {\tt Population} module, the latter having been developed to simulate the Galactic foreground signal corresponding to a given DWD population synthesis catalogue, Galactic or otherwise (as demonstrated in \citet{rieck_a_2024}). This module's approach to simulating the foreground spectrum from a DWD catalogue is detailed in \S\ref{sec:sims_pop_mw}. 

The \popmap spatial template is produced as follows: all unresolved\footnote{i.e., those with $\mathrm{SNR}<7$ with respect to a fiducial noise curve; see \S\ref{sec:sims_pop_mw}.} DWDs in the population synthesis catalogue are binned by their sky location into the pixels of a Healpy skymap in the SSB frame. The unnormalized amplitude of each pixel is determined by summing the distance-modulated GW contributions of all DWDs contained therein, and the entire skymap is then normalized to 1 per Eq.~\eqref{eq:blip_Pofn_norm}. This produces a template corresponding to a single realization of a population synthesis catalogue.

\subsection{Foreground Spectral Model}\label{sec:foreground_spec}
To model the foreground spectral distribution, we employ a tanh-truncated power law, of the kind often used in the literature to model the Galactic foreground spectrum and its turn-off around a few mHz \citep[e.g.,][]{robson_the_2019}. It is described by four parameters: the amplitude $\Omega_{\mathrm{ref}}$, the power law slope $\alpha$, the cutoff frequency $f_{\mathrm{cut}}$, and the cutoff frequency scaling parameter $f_{\mathrm{scale}}$, such that
\begin{equation}\label{eq:blip_tpl}
    \Omega_{\mathrm{GW}}(f) = \frac{1}{2}\,\Omega_{\mathrm{ref}}\left(\frac{f}{f_{\mathrm{ref}}}\right)^{\alpha} \bigg(1 + \tanh\left(\frac{f_{\mathrm{cut}} - f}{f_{\mathrm{scale}}}\right)\bigg).
\end{equation}
This formulation neglects some subtle effects on the foreground spectrum induced by iterative subtraction of resolved DWD signals \citep[see e.g.,][]{robson_the_2019} and assumes a smoother foreground spectrum than is likely in reality. However, as this study focuses on spatial characterization of the Galactic foreground, we leave further development of refined spectral models to future work.

\subsection{LISA Instrumental Noise Model}\label{sec:noise}
We employ a simple model of the LISA detector noise contributions: a combination of position and acceleration noise spectra ($S_p$ and $S_a$, respectively) as given in \citet{amaro-seoane_laser_2017}:
\begin{equation}\label{eq:blip_noise_functions}
    \begin{split}
        S_p(f) &= N_p\left[1+\left(\frac{2\,\mathrm{mHz}}{f}\right)^4\right]\,\mathrm{Hz}^{-1},\\
        S_a(f) &= \left[1+\bigg(\frac{0.4\,\mathrm{mHz}}{f}\bigg)^2\right]\,\\
        &\qquad\quad\quad\,\,\,\times\left[1+\bigg(\frac{f}{8\,\mathrm{mHz}}\bigg)^4\right] \frac{N_a}{(2\pi f)^4}\,\mathrm{Hz}^{-1}.
    \end{split}
\end{equation}
\ed{All analyses in this work employ log-uniform priors on $N_p$ and $N_a$, such that $\pi(\log_{10}N_p) = \mathcal{U}(-44,-39)$ and $\pi(\log_{10}N_a) = \mathcal{U}(-51,-46)$.} For further details on the LISA noise model in \blip, refer to \citet{banagiri_mapping_2021}. It is worth noting that LISA noise is expected to be significantly more complex in reality; as such the precision of the spectral parameter posteriors presented in \S\ref{sec:results} should not be taken as representative of what we expect to see when LISA flies. More sophisticated treatments of the LISA instrumental noise are in development \citep[e.g.,][and discussions therein]{littenberg_prototype_2023,bayle_unified_2023, hartwig_stochastic_2023, pagone_noise_2024} and can be incorporated in future.

\subsection{Simulations}\label{sec:sims}
\blip can simulate time-domain LISA data consisting of LISA instrumental noise and SGWB contributions. It does so under the assumption of rigid-body orbits for the LISA constellation. Anisotropies are incorporated by convolving a simulated skymap with the full time-dependent LISA response in segments of duration $10^5$ s under an adiabatic approximation for the LISA satellite motion. For further details of \blip simulation routines, refer to \citet{banagiri_mapping_2021,rieck_a_2024}. All simulations are performed in the pixel basis. We simulate two datasets, each spanning the full nominal LISA mission duration of 4 years.

\begin{figure*}
    \centering
    \includegraphics[width=0.8\textwidth]{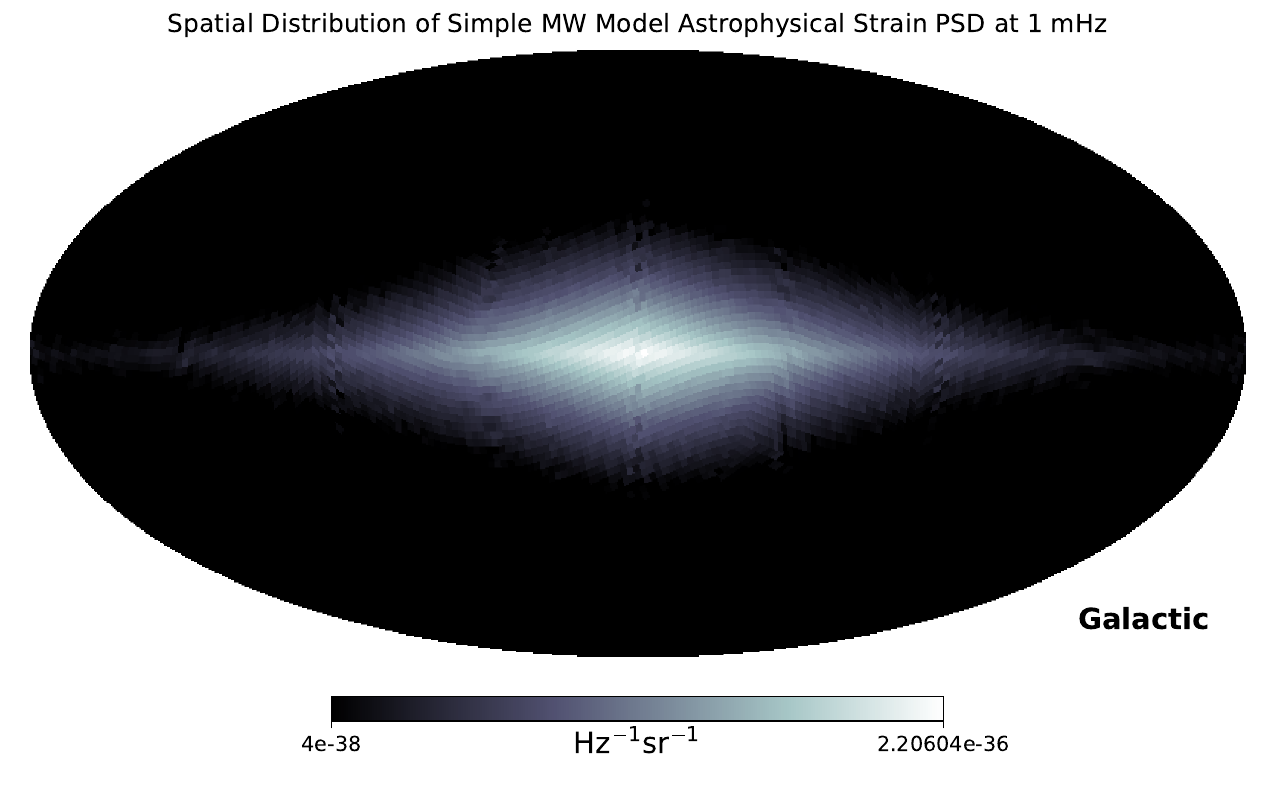}
    \caption{Spatial distribution of the simple MW foreground simulation's astrophysical strain PSD at 1 mHz, displayed with a logarithmic color scale. Pixels with zero power are shown in black. The simulation uses a Healpix {\tt nside} of 32.}
    \label{fig:simple_mw_simulated_skymap}
\end{figure*}

\begin{figure*}
    \centering
    \includegraphics[width=0.8\textwidth]{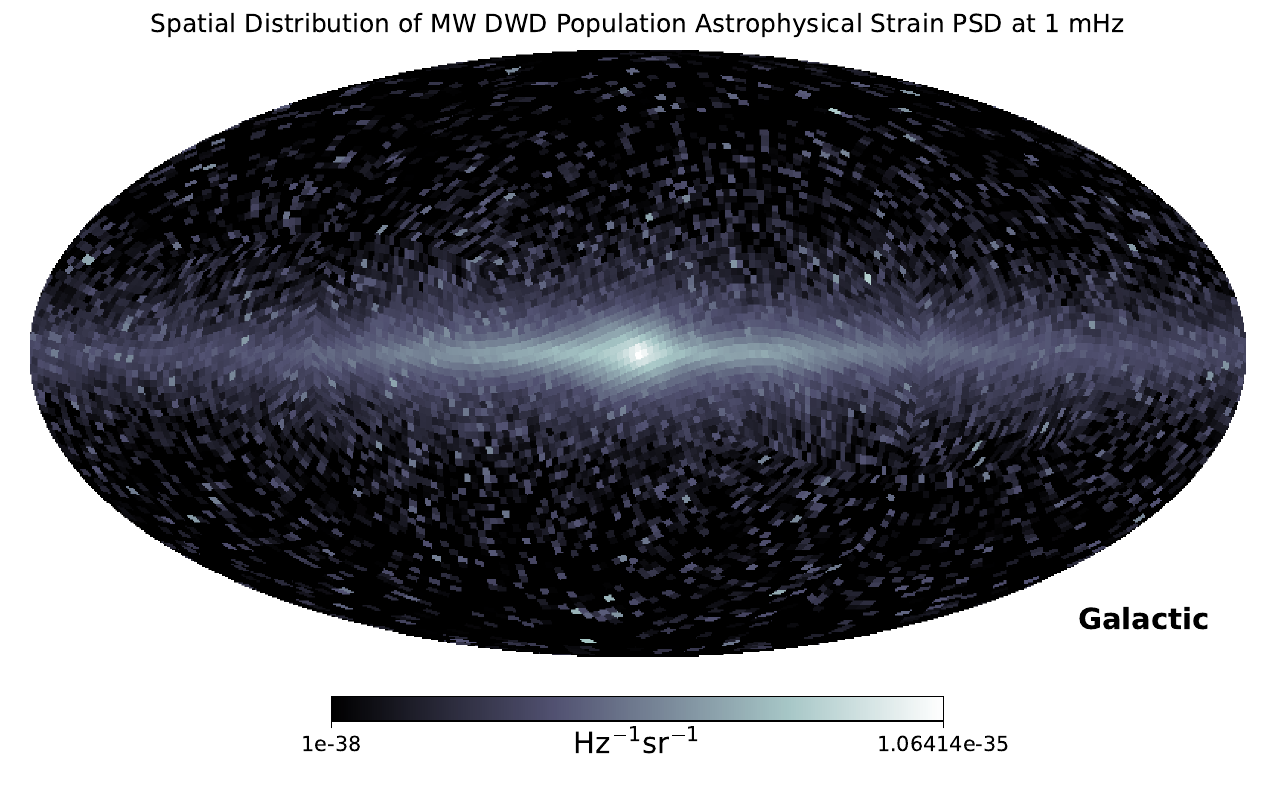}
    \caption{Spatial distribution of the \citet{wilhelm_the_2020} unresolved MW DWD population's astrophysical strain PSD at 1 mHz, displayed with a logarithmic color scale. Pixels with zero power are shown in black. The simulation uses a Healpix {\tt nside} of 32.}
    \label{fig:blip_mw_population_skymap}
\end{figure*}

\subsubsection{Simple MW Simulation}\label{sec:sims_simple_mw}
We first consider a simple MW model, with the parametric spatial distribution described in \S\ref{sec:template_simple}. We employ the thick disk model described in \citet{breivik_constraining_2020}, with $z_h=0.9$ kpc and $r_h=3.31$ kpc. The projected spatial distribution of this simple MW as seen by LISA is shown in Fig.~\ref{fig:simple_mw_simulated_skymap}. The spectral distribution of this simple MW foreground simulation is described by the tanh-truncated power law given in \S\ref{sec:foreground_spec}, with $\Omega_{\mathrm{ref}}=1.87\times10^{-6}$, $\alpha=0.667$, $f_{\mathrm{cut}}=2.08$ mHz, and $f_{\mathrm{scale}} = 1.18$ mHz. These parameters were chosen such that the overall simulated spectral shape and amplitude of this simple model are in line with expectations of the Galactic foreground spectrum.

\subsubsection{MW Population Simulation}\label{sec:sims_pop_mw}
We consider also the more realistic case of the MW foreground arising from a simulated Galactic DWD population. This is accomplished through \blips {\tt Population} module, which is designed for rapid computation of the SGWB spectrum and skymap arising from catalogues of DWDs produced by population synthesis simulations. The DWD SGWB PSD is estimated by computing the individual (assumed monochromatic) PSD contribution of each catalogue binary, as well as its fiducial SNR with respect to the \citet{robson_the_2019} LISA noise curve, including a fiducial Galactic foreground noise contribution. DWDs with SNR $>7$ are discarded as resolved, and the remainder are binned at the \blip simulation frequency resolution.\footnote{This resolution corresponds to the splice duration of \blips data generation procedure; see \citet{banagiri_mapping_2021} for details.} The skymap is produced by a binning procedure across a Healpix skymap, and normalized such that the skymap integrates to 1 across the sky per Eq.~\eqref{eq:blip_Pofn_norm} (as described in \S\ref{sec:template_pop}). The spectrum and anisotropic spatial distribution is then passed to \blips data simulation routines as normal --- i.e., a series of inverse fast Fourier transforms is applied to the spectrum to create short, 50\%-overlapping time-domain data segments, which are then spliced together; see \citet{banagiri_mapping_2021} for further details on \blips data generation procedure. When the resulting data is returned to the frequency domain via Fourier transform, this procedure produces a fast and accurate estimate of the DWD PSD. However, it is worth noting that the spectrum is slightly smoothed in the process, and as such this method may not be well-suited to investigating sharp features of the MW foreground spectrum like that of cataclysmic variables \citep{scaringi_cataclysmic_2023}.  A full-fidelity solution is, of course, known; one can simply simulate some $20-30$ million DWD waveforms and create an accurate time-domain datastream, which could then be directly added to the \blip time-domain data produced from noise and other signals with functional spectral forms. Such an approach is, however, quite computationally expensive, although recent advances such as GBGPU \citep{katz_mikekatz04/gbgpu_2024} have made doing so significantly more approachable. Implementing a full time-domain approach to computing unresolved DWD spectra in \blip is a promising direction of future work. That being said, the ability to produce rapid, high-fidelity simulations of the MW foreground spatial distribution from DWD population synthesis catalogues carries significant utility for this work.

We apply this method to the population synthesis catalogue of \citet{wilhelm_the_2020}. The \citet{wilhelm_the_2020} MW simulation combines the detailed DWD binary population synthesis model of \citet{toonen_supernova_2012} with a high-resolution simulation of Galactic dynamics \citep{donghia_trojans_2020} to create a MW DWD population that follows a highly realistic spatial distribution in the MW, including a complex bar and spiral arm structure. For further details on this DWD population, see \citet{wilhelm_the_2020}. The spatial distribution of this population's anisotropic Galactic foreground is shown in Fig.~\ref{fig:blip_mw_population_skymap}.

\section{Results}\label{sec:results}
We present two demonstrations of the ability of our templated anisotropic analysis to characterize the Galactic foreground. \ed{We additionally consider the impacts of a poor choice of template on recovery of the foreground spectrum.} All analyses discussed are performed in the pixel basis with a Healpix {\tt nside} (pixel resolution) of 32. This resolution was chosen due to computational limitations in the simulation process; ASGWB simulations with finer pixel resolutions will be accessible in future through further optimization of \blips simulation routines and/or the availability of additional computational resources. For each case considered, we use Hamiltonian Monte Carlo to sample its respective posterior distribution via the Numpyro package \citep{phan_composable_2019a}. Each analysis was performed on a single Nvidia A100 GPU.

\clearpage

\subsection{Templated Recovery of a Simple Galactic Foreground}\label{sec:results_simple}
First, we consider the simple MW foregound simulation discussed in \S\ref{sec:sims_simple_mw}. We recover the signal in the presence of LISA instrumental noise as described in \S\ref{sec:noise}, with the tanh-truncated power law spectral model of Eq.~\eqref{eq:blip_tpl}. We assume that the low-frequency slope $\alpha=2/3$ --- as expected for a compact binary SGWB --- and infer the remaining parameters. \ed{This choice is made due to the presence of a degeneracy between $\alpha$ and $f_{\mathrm{scale}}$ in this spectral model. Namely, the spectrum truncation can mimic a different slope within LISA's most-sensitive frequency range. For the idealized scenario considered here, this can result in biases in parameter recovery, as small variations in $\alpha$ and $f_{\mathrm{scale}}$ from the ``true" (simulated) parameter values are able to accurately fit the most-informative data. Fixing $\alpha$ breaks this degeneracy and allows this simple case to serve its purpose as a test of the templated anisotropic analysis.\footnote{It is worth noting that this degeneracy may impact attempts to recover foreground astrophysics from its spectral characteristics while using this (common) spectral form. However, as can be seen in \S\ref{sec:results_population}, this degeneracy is less relevant for the realistic case where the simulation and model spectra are not identical in form.} Given the focus of this work on spatial analyses, we note the development of improved foreground spectral models as an important direction of future work and carry on with this idealized test case.}

\begin{figure*}
    \centering
    \includegraphics[width=0.8\textwidth]{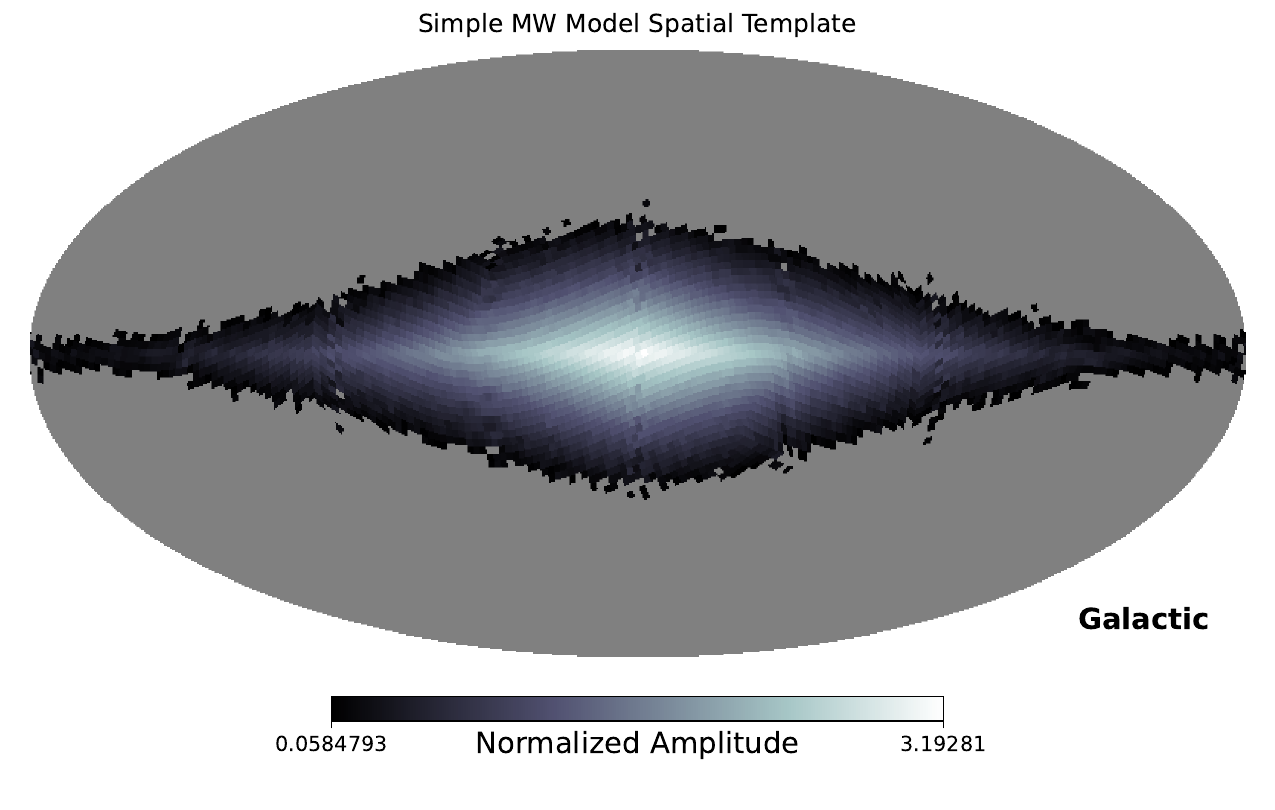}
    \caption{Simple MW model template skymap with a Healpix {\tt nside} of 32, normalized per Eq.~\eqref{eq:blip_Pofn_norm}. Pixels in grey are masked (i.e., zero-amplitude).}
    \label{fig:blip_simple_mw_fixedsky_map}
\end{figure*}

\begin{figure*}
    \centering
    \includegraphics[width=0.75\textwidth]{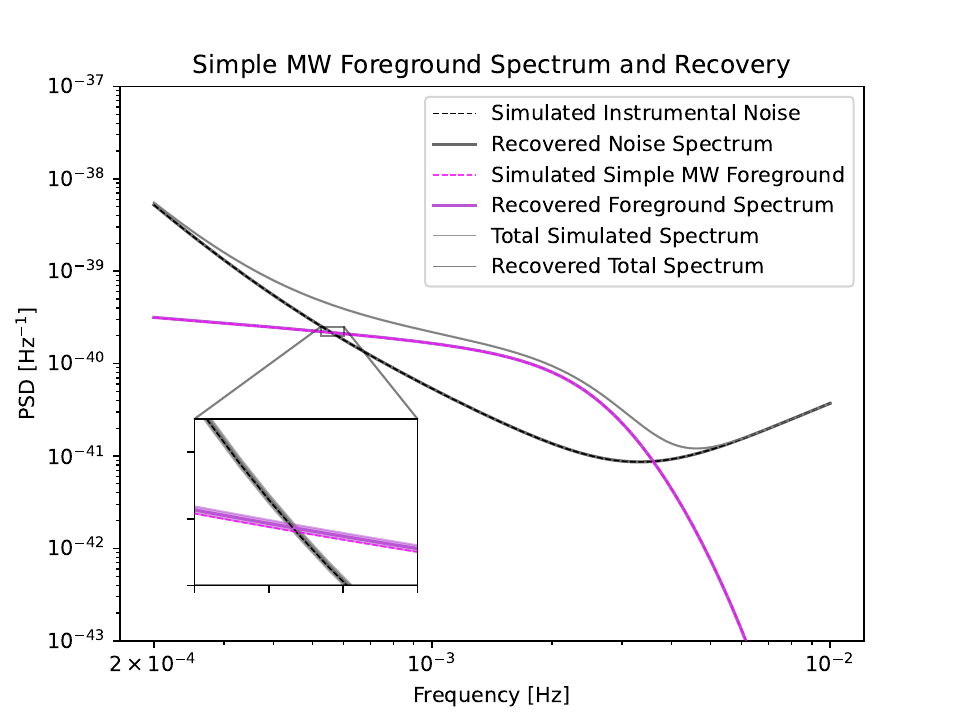}
    \caption{Simulated and recovered PSDs of both the LISA instrumental noise and the simple MW foreground. The simulated spectra are shown as dashed lines, the recovered spectral medians as solid lines, and the 95\% C.I. bounds are shown as shaded regions. As all spectral parameters are precisely recovered, the 95\% C.I. lie extremely close to the simulated/median spectra and are therefore difficult to distinguish by eye; an inset has been provided to give the reader a sense of the 95\% C.I. width. As LISA's instrumental noise is expected to be significantly more complex than is considered here, we do not expect this precision to be representative of what LISA will achieve when it flies.}
    \label{fig:blip_simple_mw_fixedsky_spec}
\end{figure*}

\begin{figure*}
    \centering
    \includegraphics[width=0.9\textwidth]{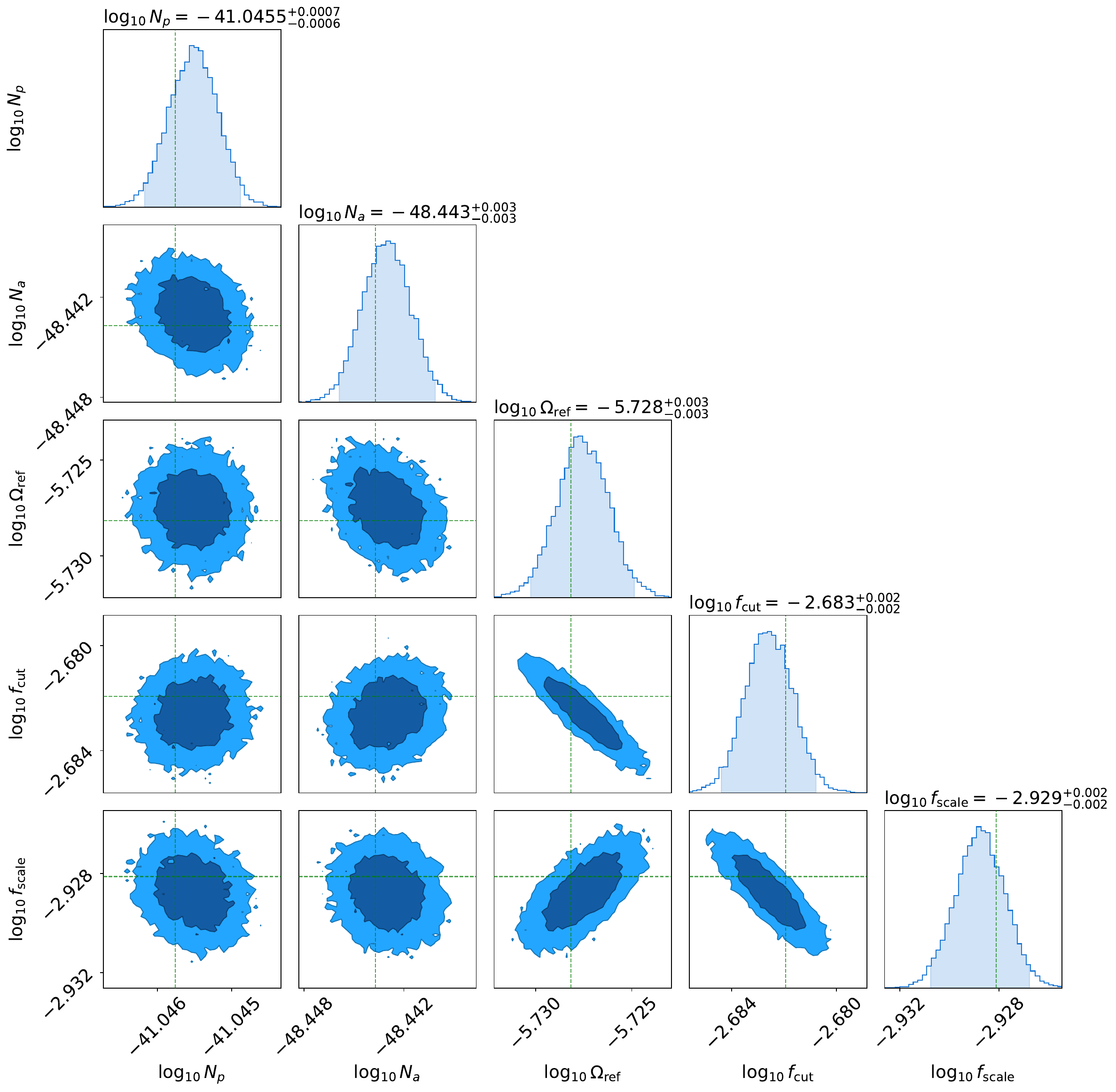}
    \caption{Corner plot of posterior samples for the simple MW model analysis. The simulated parameter values are marked by green lines. Quoted bounds are the mean and 95\% C.I. of the posterior samples. The shading in the one-dimensional posterior distribution denotes the 95\% C.I.; the dark and light shaded regions of the two-dimensional distributions denote $1$- and $2\sigma$ bounds, respectively. All parameters are recovered precisely and without bias.}
    \label{fig:blip_simple_mw_fixedsky_corner}
\end{figure*}

We employ uniform priors with astrophysically-motivated bounds, such that $\pi(\log\Omega_{\mathrm{ref}}) = \mathcal{U}(-6,-3)$, $\pi(\log_{10}f_{\mathrm{cut}}) = \mathcal{U}(-3.1,-2.4)$, and $\pi(\log_{10}f_{\mathrm{scale}}) = \mathcal{U}(-4,-2)$.  We assume that the parameters of the simple MW spatial model have been well-measured by analyses of the resolved DWD systems, and perform a templated analysis with a spatial distribution given by the \bgal model of \S\ref{sec:template_simple} with $r_h = 3.31$ kpc and $z_h=0.9$ kpc. The assumed spatial template is shown in Fig.~\ref{fig:blip_simple_mw_fixedsky_map} and the simulated and recovered spectra are shown in Fig.~\ref{fig:blip_simple_mw_fixedsky_spec}. The noise and foreground spectral parameters are recovered without bias, as can be seen in the sampled posterior distributions shown in Fig.~\ref{fig:blip_simple_mw_fixedsky_corner}.

\subsection{Templated Recovery of a Realistic, Population-Derived Galactic Foreground}\label{sec:results_population}
Having validated our method on a simple model with known spectral parameters, we will now consider the more complex case of a realistic, population-derived simulated MW foreground as described in \S\ref{sec:sims_pop_mw}.

\begin{figure*}
    \centering
    \includegraphics[width=0.8\textwidth]{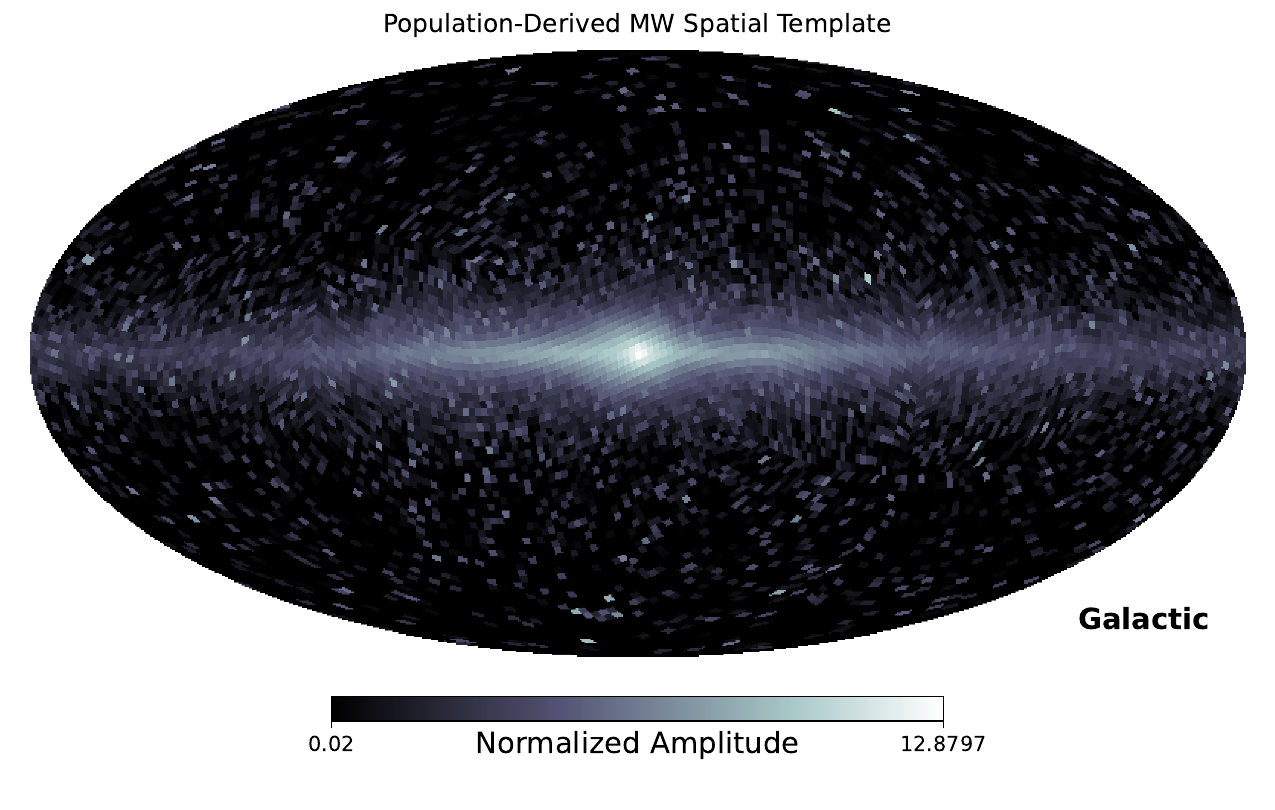}
    \caption{Population-derived MW foreground template skymap with a Healpix {\tt nside} of 32, normalized per Eq.~\eqref{eq:blip_Pofn_norm}.}
    \label{fig:blip_mw_popsky_map}
\end{figure*}

\begin{figure*}
    \centering
    \includegraphics[width=0.75\textwidth]{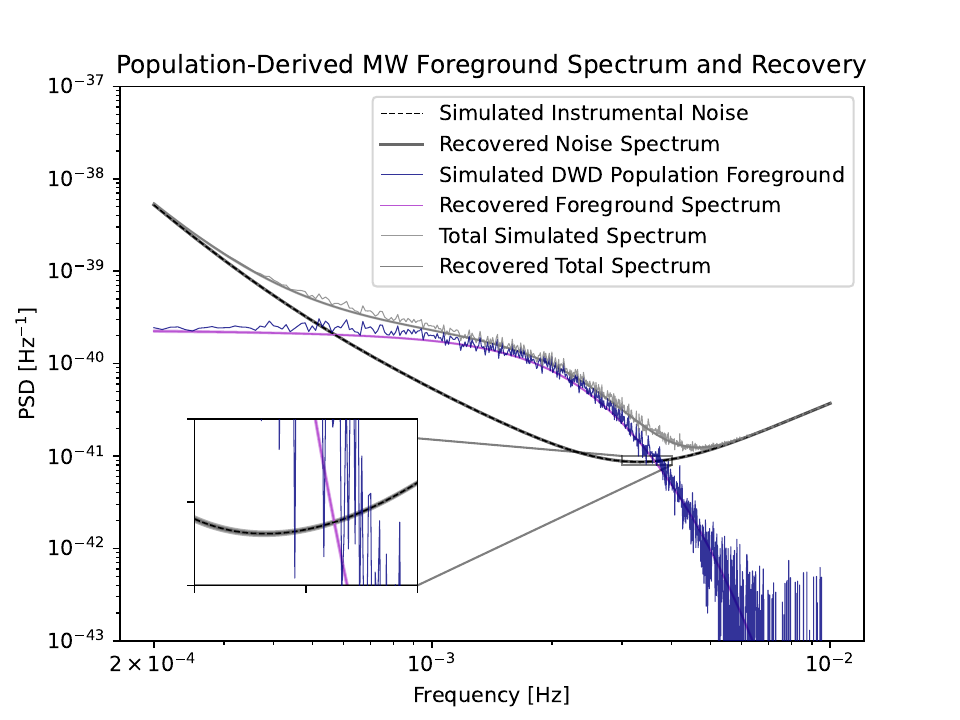}
    \caption{Simulated and recovered PSDs of both the LISA instrumental noise and the realistic, population-derived MW foreground. The simulated noise spectra are shown as dashed lines, the simulated MW spectra are given in narrow lines, the recovered spectral medians as solid lines, and the 95\% C.I. bounds are shown as shaded regions. As all spectral parameters are precisely recovered, the 95\% C.I. lie extremely close to the simulated/median spectra and are therefore difficult to distinguish by eye; an inset has been provided to give the reader a sense of the 95\% C.I. width. As LISA's instrumental noise is expected to be significantly more complex than is considered here, we do not expect this precision to be representative of what LISA will achieve when it flies.}
    \label{fig:blip_mw_popsky_spec}
\end{figure*}

\begin{figure*}
    \centering
    \includegraphics[width=0.9\textwidth]{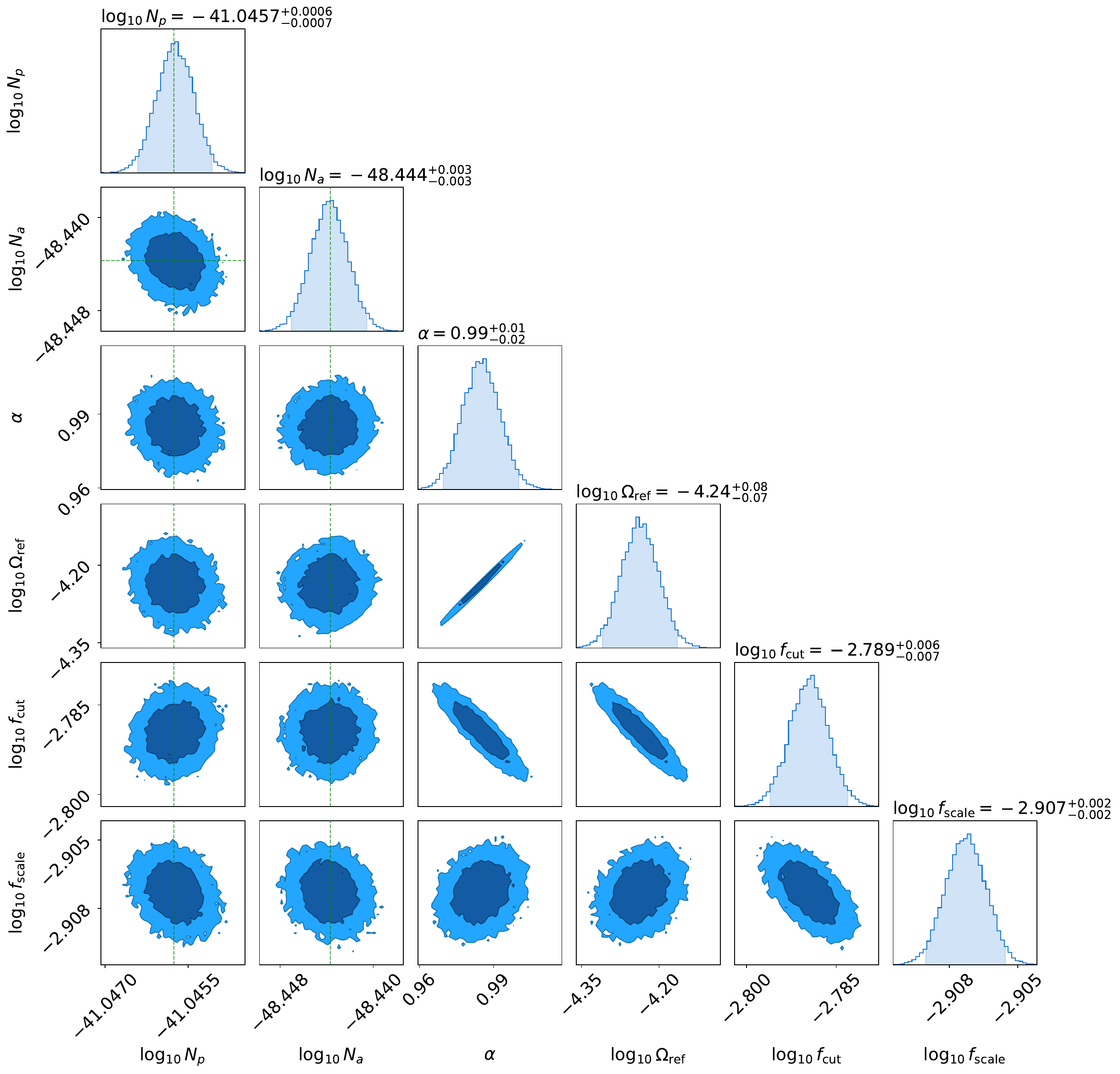}
    \caption{Corner plot of posterior samples for the realistic, population-derived MW foreground analysis. The true values of the simulated parameters are marked by green lines; no true values exist for the population-derived MW foreground. Quoted bounds are the mean and 95\% C.I. of the posterior samples. The shading in the one-dimensional posterior distribution denotes the 95\% C.I.; the dark and light shaded regions of the two-dimensional distributions denote $1$- and $2\sigma$ bounds, respectively.}
    \label{fig:blip_mw_popsky_corners}
\end{figure*}

As in the simpler case, we recover the population-derived foreground under the assumption that the distribution of the population is well-measured by independent inference of the resolved DWD population. We employ the {\tt popmap} spatial model discussed in \S\ref{sec:template_pop} with the spatial distribution shown in Fig.~\ref{fig:blip_mw_popsky_map}. We again use the tanh-truncated power law spectral model of Eq.~\eqref{eq:blip_tpl}, but additionally infer the low-frequency slope $\alpha$. The priors are again uniform with astrophysically-motivated bounds, such that $\pi(\log\Omega_{\mathrm{ref}}) = \mathcal{U}(-6,-3)$, $\pi(\alpha) = \mathcal{U}(0,2)$, $\pi(\log_{10}f_{\mathrm{cut}}) = \mathcal{U}(-3.1,-2.4)$, and $\pi(\log_{10}f_{\mathrm{scale}}) = \mathcal{U}(-4,-2)$. The simulated and recovered spectra are shown in Fig.~\ref{fig:blip_mw_popsky_spec}, the posterior distributions for each parameter are shown in Fig.~\ref{fig:blip_mw_popsky_corners}. Using this templated approach, we are able to accurately characterize the overall MW spectral distribution without bias, although we note again that more complex spectral models should be explored in future to account for the bin-to-bin variance of the foreground amplitude. This analysis, performed on 4 years of LISA data, took less than 1 day on a single Nvidia A100 GPU.

\begin{figure*}
    \centering
    \includegraphics[width=0.8\textwidth]{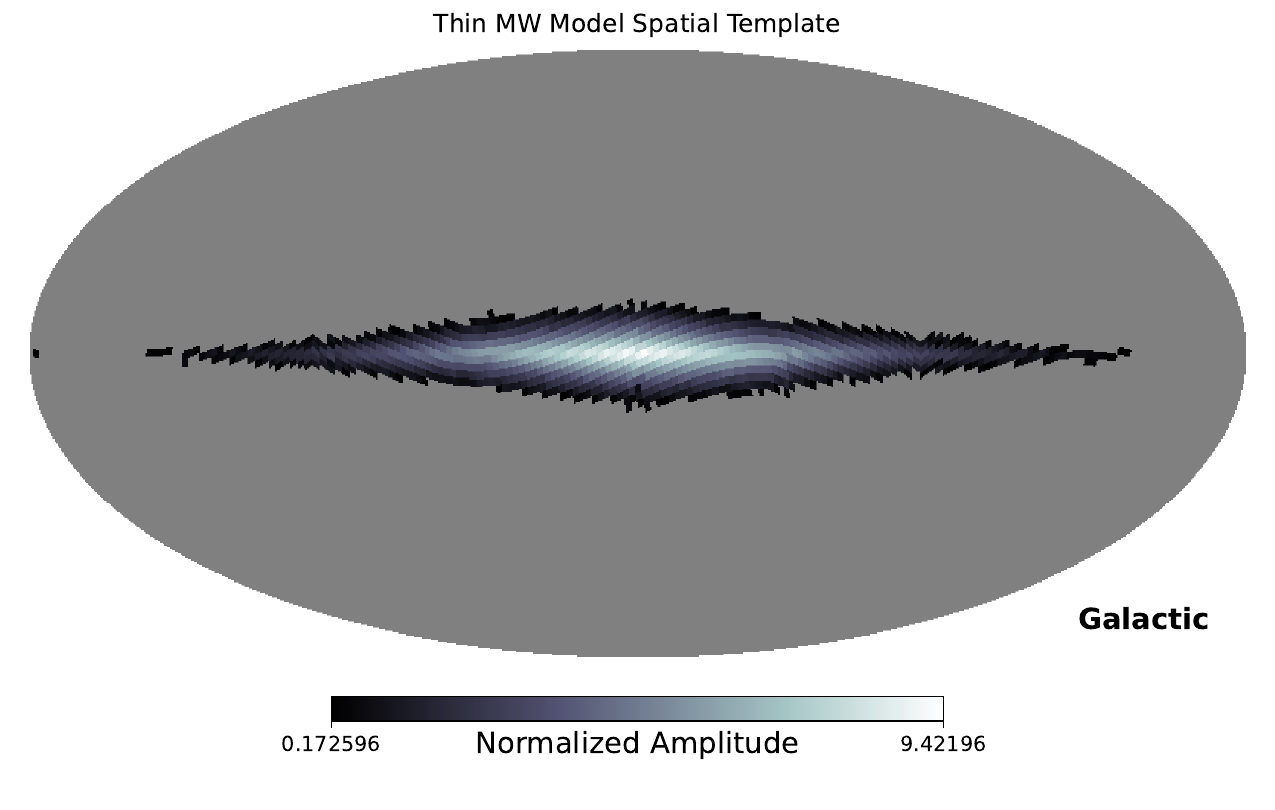}
    \caption{\ed{The simple thin-disc MW model template skymap used for the mismodelling analysis discussed in \S\ref{sec:results_mismatch}. The template uses a Healpix {\tt nside} of 32, and is normalized per Eq.~\eqref{eq:blip_Pofn_norm}. Pixels in grey are masked (i.e., zero-amplitude).}}
    \label{fig:blip_mismatch_map}
\end{figure*}

\begin{figure*}
    \centering
    \includegraphics[width=0.75\textwidth]{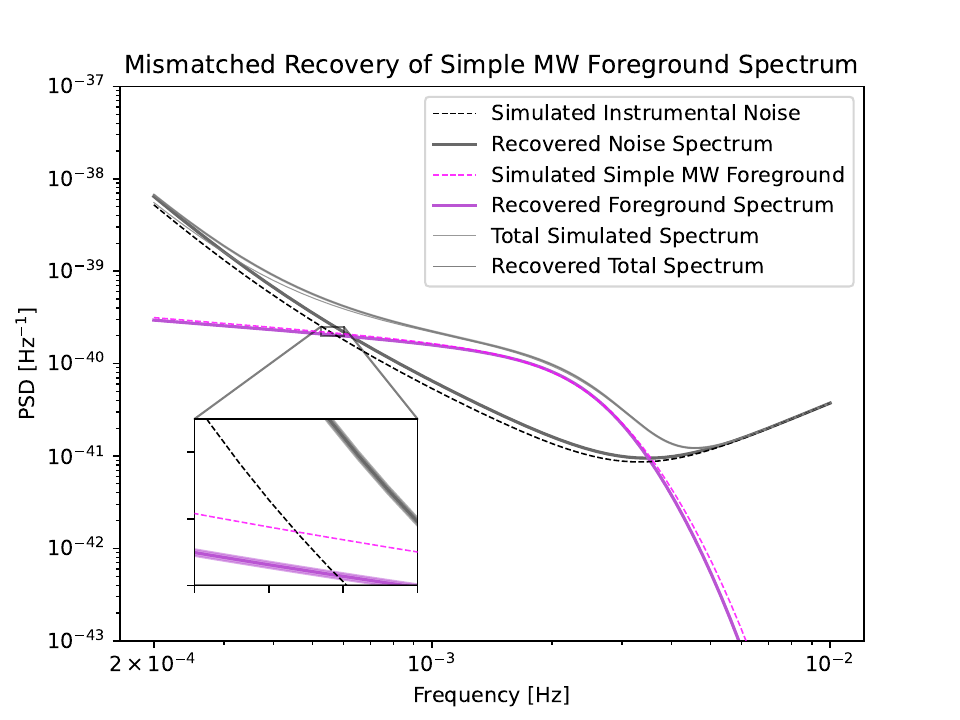}
    \caption{\ed{Simulated and recovered PSDs of both the LISA instrumental noise and the simple MW foreground for the mismodelling analysis of \S\ref{sec:results_mismatch}. The simulated spectra are shown as dashed lines, the recovered spectral medians as solid lines, and the 95\% C.I. bounds are shown as shaded regions. The recoveries of both the LISA instrumental noise and the foreground spectrum, while nonetheless precise, exhibit a significant systematic bias due to mismodelling of the foreground anisotropy.}}
    \label{fig:blip_mismatch_spec}
\end{figure*}

\begin{figure*}
    \centering
    \includegraphics[width=0.9\textwidth]{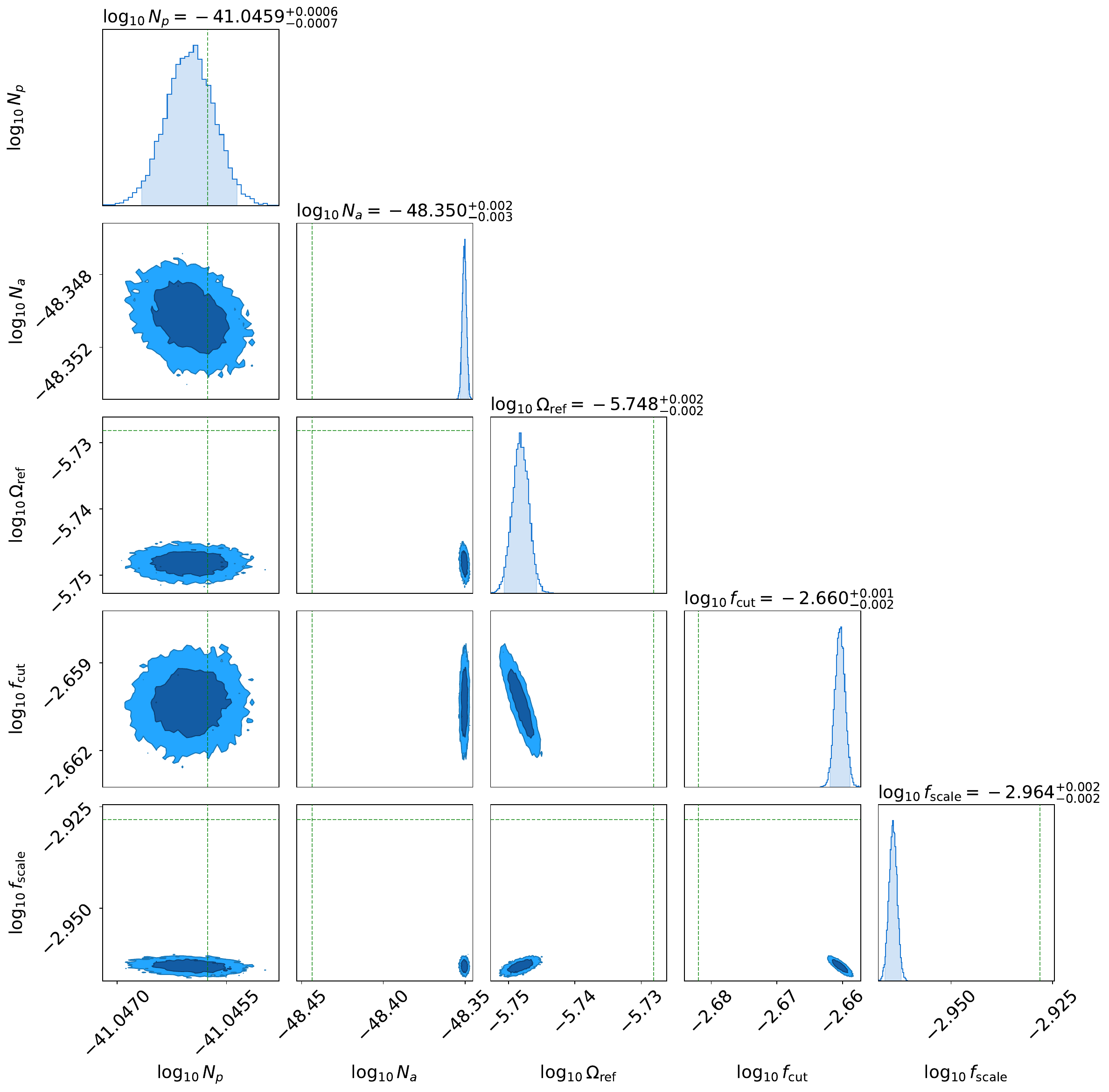}
    \caption{\ed{Corner plot of posterior samples for the spatial mismodelling analysis discussed in \S\ref{sec:results_mismatch}. The simulated parameter values are marked by green lines. Quoted bounds are the mean and 95\% C.I. of the posterior samples. The shading in the one-dimensional posterior distribution denotes the 95\% C.I.; the dark and light shaded regions of the two-dimensional distributions denote $1$- and $2\sigma$ bounds, respectively. All the foreground spectral parameters' posteriors are severely biased due to mismodelling of the foreground anisotropy, as is the acceleration noise parameter $N_a$. The position noise is recovered without bias, due to its dominance at high frequencies (where the foreground contribution is minimal).}}
    \label{fig:blip_mismatch_corners}
\end{figure*}

\ed{
\subsection{Effects of Spatial Mismodelling}\label{sec:results_mismatch}
Finally, we investigate the impact of spatial mismoddelling on the foreground spectral recovery. We consider again the simple MW simulation of \ref{sec:sims_simple_mw} and repeat the analysis detailed in \ref{sec:results_simple} with one crucial alteration: we assume a ``thin disc" spatial template, given by the \bgal spatial model with $r_h = 2.9$ kpc and $z_h = 0.3$ kpc. This template is shown in Fig.~\ref{fig:blip_mismatch_map}. Despite this, as can be seen in the recovered spectra (Fig.~\ref{fig:blip_mismatch_spec}) and individual parameter recoveries (Fig.~\ref{fig:blip_mismatch_corners}), the spatial mismodelling induces biases in both the individual parameters, as well as the overall spectral recovery. It is worth noting that this is a fairly conservative example, as we assume an identical form for the spatial distribution and only alter its parameters. This result indicates a need for careful model choice in anisotropic analyses of the Galactic foreground. That being said, it also indicates the level to which the data can be informative as to different spatial models --- thereby lending additional motivation for fully-parameterized astrophysical models of the foreground anisotropy.}

\section{Discussion and Future Work}\label{sec:conclusion}
We present a novel analysis which performs a targeted Bayesian search for the LISA Galactic DWD foreground via pixel-basis anisotropic templates. We demonstrate this method on 4 years of simulated LISA data and show that we are able to use our templated anisotropic analysis to efficiently and accurately recover the Galactic foreground spectrum for both a simplistic model of the Milky Way and a more complex, DWD population-derived realization of the foreground. Not only is this templated approach $\sim10-100\times$ more computationally efficient than previous analyses of the LISA Galactic foreground via a spherical harmonic spatial model \citep[as demonstrated in][]{banagiri_mapping_2021}, but it is also able to more accurately model the spatial distribution of the Galactic foreground through leveraging information gleaned from its counterpart resolved DWDs. To the authors' knowledge at time of writing, this work comprises the first demonstration of an explicitly anisotropic recovery of the LISA Galactic foreground driven by the underlying astrophysical spatial distribution of its contingent binary systems.

There remain, however, several promising directions of future work. It will be important to \ed{further} characterize the effects of mismodelling the foreground anisotropy on inference of its spectral distribution. \ed{As we show in \S\ref{sec:results_mismatch}, incorrect assumptions as to the foreground anisotropy} induce biases in recovery of the foreground spectrum. \ed{As such,} robust spatial inference methods for the Galactic foreground take on additional importance for the success of the LISA mission as a whole. Conversely, \ed{this result} indicates substantial constraining power for astrophysically-parameterized spatial models of the foreground anisotropy. \ed{Of particular interest for future investigation is the impact of not accounting for Galactic structures such as a bar, different spiral arm configurations, or even the presence of unresolved DWDs in previously unknown satellite galaxies behind the Galactic disc as discussed in, e.g., \citep{pozzoli_cyclostationary_2024}.} Investigations into \ed{anisotropy-induced} spectral biases of this kind will also depend in part on development of refined spectral models for the both the LISA instrumental noise and the foreground spectrum itself. In particular, it will be important to determine the effects of non-stationary instrumental noise on the recovery of anisotropic signals, as the latter is dependent on the specific time-dependence of the LISA detector response to an anisotropic signal.

It is worth noting that the the inherent Poisson variance of a single population synthesis simulation may hinder the efficacy of the \popmap template under relaxed assumptions as to our knowledge of the MW foreground anisotropy. A possible extension of this approach could resolve or reduce the impact of this consideration by repeating the process described in \S\ref{sec:template_pop} for a large set of catalogues with varied underlying assumptions that span our prior uncertainty as to the MW DWD population, and creating a single template that averages over all of them.\footnote{As was considered in \citet{agarwal_targeted_2022} for a targeted analysis of the Galactic plane with terrestrial GW observatories.}

This work possesses immediate applications in several other areas of SGWB science with LISA. Of particular interest is the utility of this templated approach towards foreground characterization when applied to the problem of spectral separation between the Galactic foreground and underlying SGWBs (e.g., the extragalactic stellar-origin binary SGWB and/or the ASGWB from DWDs in the Large Magellanic Cloud). Additionally, analyses of the latter anisotropic signal could benefit significantly from a similar targeted search via a pixel-basis template. The spatial distribution of the Large Magellanic Cloud on the sky is well-understood even in comparison to that of the Milky Way; as such it is exceptionally well-suited to a fixed-template approach.\footnote{The $\sim100$ or so individually resolved DWD systems we expect to observe in the Large Magellanic Cloud with LISA \citep{korol_weighing_2021} are not likely to provide more robust spatial constraints on the LMC than extant electromagnetic studies. As such, the Large Magellanic Cloud anisotropy is more-or-less known \textit{a priori} for LISA's purposes.}

Finally, this advancement opens the door for development of astrophysically-parameterized models of the foreground anisotropy, either via template banks or continuous parameterized models. These approaches would allow for direct inference of parameters describing the 3D spatial distribution of the MW DWD population, allowing for a unique insight into the MW as it existed when these systems first formed. Additionally, there are immediate applications for such models in the LISA Global Fit setting; spatial inference of the resolved DWD population \citep[as demonstrated in e.g., ][]{littenberg_prototype_2023} could then be immediately liaised into improved modelling of the foreground anisotropy. Such an approach could easily comprise a joint population inference model that considers both resolved and unresolved LISA DWDs, leveraging all available information to infer the overall spatial distribution of this compelling astrophysical population.

\begin{acknowledgments}
This manuscript was adapted in part from the doctoral dissertation of AWC, \textit{Astrophysical Inferences from Multimessenger Ensembles}. The work described was supported by NASA grant 90NSSC19K0318 and NSF grant no. 2125764. The authors acknowledge the use of computing resources provided by the Minnesota Supercomputing Institute at the University of Minnesota. Packages used in this work include {\tt numpy} \citep{harris_array_2020a}, {\tt scipy} \citep{virtanen_scipy_2020a}, {\tt ChainConsumer} \citep{hinton_chainconsumer_2016}, {\tt astropy} \citep{robitaille_astropy:_2013,theastropycollaboration_the_2018,astropycollaboration_the_2022a}, JAX \citep{bradbury_jax:_2018a}, Numpyro \citep{phan_composable_2019a},and {\tt Matplotlib} \citep{hunter_matplotlib:_2007}. \ed{The authors thank the anonymous reviewer for detailed comments which improved the manuscript as a whole.} We extend our gratitude to Sharan Banagiri, Joe Romano, Jessica Lawrence, and Malachy Bloom for their work on \blip and many helpful conversations.
\end{acknowledgments}

\section*{Data Availability}
\blip is open source; the primary repo can be found at \hyperlink{https://github.com/sharanbngr/blip}{https://github.com/sharanbngr/blip}. The specific version of the code used in this study can be found within the corresponding BLIP release on Zenodo \citep{criswell_criswellalexander/blip:_2024}.  Simulated data and posterior samples for all analyses are available on Zenodo \citep{criswell_datasets_2024}.
\clearpage

\bibliography{mw_templates.bib} 
\bibliographystyle{apsrev4-2.bst}

\end{document}